\documentstyle[12pt]{article}
\def\ba{\begin{eqnarray}}
\def\ea{\end{eqnarray}}
\def\nn{\nonumber}
\def\l{\label}
\def\p{\phi}
\begin{document}
\setlength{\unitlength}{1mm}
{\hfill  JINR E2-95-280, June 1995; hep-th/9506206} \vspace*{2cm} \\
\begin{center}
{\Large\bf
Two-dimensional Quantum-Corrected Eternal Black Hole}
\end{center}
\begin{center}
{\large\bf  Sergey N.~Solodukhin$^{\ast}$}
\end{center}
\begin{center}
{\bf Bogoliubov Laboratory of Theoretical Physics, Joint Institute for
Nuclear Research, Head Post Office, P.O.Box 79, Moscow, Russia}
\end{center}
\vspace*{2cm}
\begin{abstract}
The one-loop quantum corrections to geometry and
thermodynamics of black hole are studied for the two-dimensional RST model.
We chose boundary conditions corresponding to the eternal black
hole being in the thermal equilibrium with the Hawking radiation.
The equations of motion are exactly integrated.
 The one of the solutions obtained is the constant curvature
 space-time with dilaton being a constant function. Such a
 solution is absent in the classical theory. On the other hand,
  we derive
the quantum-corrected metric (\ref{solution}) written in the
Schwarzschild like form which is a deformation of the classical
black hole solution \cite{5d}. The space-time singularity occurs
to be milder than in classics and the solution admits two
asymptotically flat black hole space-times lying at "different sides"
of the singularity. The thermodynamics of the classical black hole
and its quantum counterpart is formulated. The thermodynamical
quantities (energy, temperature, entropy) are calculated and occur
to be the same for both the classical and quantum-corrected black holes.
So, no quantum corrections to thermodynamics are observed.
The possible relevance of the results obtained to the four-dimensional
case is discussed.
\end{abstract}
\begin{center}
{\it PACS number(s): 04.60.+n, 12.25.+e, 97.60.Lf, 11.10.Gh}
\end{center}
\vskip 1cm
\noindent $^{ \ast}$ e-mail: solod@thsun1.jinr.dubna.su
\newpage
\baselineskip=.8cm
\section{Introduction}
\setcounter{equation}0
Interest in the quantum corrections in the gravitational theory is two-fold.
At first, it is commonly believed that a successful quantization of gravity
will provide us with modifications of the theory which are necessary
to avoid  space-time singularities typically predicted by a classical theory
of gravity \cite{1d}. These singularities occur in the Universe and inside
black holes
under rather general assumptions about properties of the matter and manifest
themselves
in the unlimited increase in the curvature of the space-time. The classical
theory is not applicable near a singularity and, in particular, we cannot
believe in its predictions concerning the complete global structure of the
space-time.
Quantum corrections may completely change the gravitational equations and the
corresponding geometry on the Planck scale and drastically modify the classical
picture \cite{2d}. The main problem on this way is the nonrenormalizability
of the Einstein gravity since the straightforward exploiting of the standard
perturbation methods leads to an inconsistent quantum theory. However, as a
first step, we can consider  the semiclassical picture when only matter fields
are quantized whereas the gravitational degrees of freedom are treated
classically.
Quantum matter fields, being integrated out in the functional integral,
 induce a term additional  to the Einstein's one in the action. The extremum of
the complete
 effective action gives us a quantum-corrected solution. Unfortunately,
 in four dimensions the effective action is not known exactly though
 it can be derived by a nonlocal polynomial with respect to
 curvatures \cite{3d}. The situation is more  hopeful in two dimensions
 where  for the conformal matter the effective action is given by the
well-known Polyakov-Liouville
 term. This was the reason why for the last years the two-dimensional
models of gravity have intensively been studied \cite{4d}.
That the theory predicts the existence of two-dimensional black holes was
stated in \cite{4}.
Then the black hole type solutions were discovered in the so-called
"string-inspired"
two-dimensional dilaton gravity \cite{5d}. It was believed  that in
two-dimensional
toy models  one could resolve the old problems of the black hole evaporation
\cite{6d} by reducing them to  solving  differential equations
of the semiclassical theory. However, the original model \cite{6d} occurred
to be not exactly integrable, which resulted in searching and
formulating a number of  exactly solvable models \cite{7d}-\cite{9d}.
Russo, Susskind and Thorlacius (RST) \cite{9d} modified the semiclassical
action by a local counterterm with which the theory becomes exactly
soluble. This RST model found a wide popularity in the context of different
aspects
of the black hole evaporation \cite{10d} and black hole thermodynamics
\cite{11d}-\cite{13d}.

The study of two-dimensional models becomes more exciting due to that the
four-dimensional Einstein theory in the spherically symmetric case reduces to
an
effective two-dimensional theory of the dilaton type \cite{14d}, \cite{15d}.
This allows one to find the effective action in the spherically symmetric
case and the corresponding quantum deformation of the classical (Schwarzschild)
configuration \cite{15d}.

\bigskip

The other point where the quantum corrections may be important is the
thermodynamics of  black hole. The most intriguing problem is
dynamical explanation of the degrees of freedom inside a hole that are counted
by the Bekenstein-Hawking formula \cite{Bekenstein1} relating the entropy of a
hole with
the area of its horizon (for the review of different approaches see
ref.\cite{Bekenstein}).
There has been much interest in this problem \cite{a1}-\cite{Barbon}
in the recent literature.
One of the ideas is that the entropy of a hole is due to quantum matter
excitations
propagating inside or just outside  the horizon. So the whole black hole
entropy
can be treated as a quantum correction. It has been shown that it is
ultraviolet
divergent \cite{a1} that can be removed by the standard renormalization
of the gravitational constant \cite{Sus}-\cite{LW}. (For the discussion of this
problem see
also refs.\cite{Kabat}, \cite{Barbon}.) Unfortunately, the classical
(tree-level)
Bekenstein-Hawking entropy does not have  dynamical explanation in this
approach.

However, in addition to divergent corrections there might be  finite
corrections to the thermodynamical quantities (mass, entropy, temperature) that
are of high interest since they may be essential at the final stages of the
black
hole evaporation when mass of the black hole becomes comparable with the
Planck mass. The corrections, logarithmically dependent on the mass of a hole,
have recently been  observed in two \cite{11d}, \cite{SS} and four \cite{F1}
dimensions
by means of the perturbative calculations on the fixed classical background.

\bigskip

The aim of this paper is to give,  within the 2D dilaton gravity modified by
the one-loop
contributions according to RST \cite{9d}, the complete and detailed
investigation
of the above-noted problems: form of the quantum-corrected geometry of
the eternal black hole and  calculation of the corresponding thermodynamical
quantities in one loop.

One remark is in order. The RST model  is exactly solvable but the solution is
uniquely defined only up to the boundary conditions that may essentially
change the character and physical interpretation of the solution. This is
because
the nonlocal nature of the Polyakov-Liouville term  the effective
action contains an ambiguity corresponding to different choices of
the quantum state of the system. The choice made in \cite{9d}  describes the
formation
of the black hole from  vacuum space-time  due to the incoming matter.
There is no any Hawking radiation for the vacuum flat space-time stage;
the radiation energy-momentum tensor is zero, $T^{rad}_{\mu\nu}=0$,
in the asymptotically flat region. Instead, we are interested in the already
formed eternal black hole being in the thermal equilibrium with the Hawking
radiation. At infinity, we have asymptotically flat space-time filled by
radiation with the energy density $T^{0,rad}_{0}={\pi \over 6}NT^2$.
Therefore, our choice of boundary conditions is different from that of
\cite{9d} to ensure this behavior at infinity.

\bigskip

Our paper is organized as follows. In Section 2. we write down the RST
equations, explain
our choice of the boundary conditions describing the eternal black hole and
find
exactly the general solution in the Schwarzschild form. The solution obtained
is a
quantum deformation of the known classical dilaton black hole \cite{5d}.
The global structure  of the found solution is studied in Section 3.
In Section 4., we give some general remarks on formulation of the black
hole thermodynamics and calculate the thermodynamical quantities (mass,
entropy,
temperature) for the classical black hole\footnotemark\footnotetext{The
thermodynamics of the classical
2D dilaton black hole has been previously studied in \cite{20}-\cite{22}.}
and   its one-loop counterpart. The comparison with the previous perturbative
calculations \cite{11d}, \cite{SS} is given. The possible relevance to the
four-dimensional
black hole physics is discussed in Section 5. The results obtained are
summarized
in the Conclusion.

\bigskip

\section{Eternal black hole solution of the RST model}
\setcounter{equation}0

{\large\bf A. Action and field equations}

The classical action of the dilaton gravity \cite{5d}
\begin{equation}
I_{0}={1 \over 2\pi} \int_{M}^{}d^2 x \sqrt{-g} e^{-2\phi}
[R+4(\nabla \phi)^2+4\lambda^2]+{1 \over \pi}\int_{\partial M}^{}e^{-2\phi}kds
\label{I0}
\end{equation}
on the quantum level, accordingly to \cite{9d}, gets modified by the following
terms:
\begin{equation}
I_{1}=-{\kappa \over 2\pi}\int_{M}^{}d^2 x \sqrt{-g} \phi R-{\kappa \over \pi}
\int_{\partial M}^{}\phi k ds
\label{I1}
\end{equation}
and
\begin{equation}
I_{2}=-{\kappa \over 2\pi} \int_{M} d^2 x \sqrt{-g}({1 \over 2}(\nabla \psi
)^2+
 \psi R) -{\kappa \over \pi}\int_{\partial M}^{}\psi k ds
\label{I2}
\end{equation}
where  we added in (\ref{I0}) and (\ref{I1}) the boundary terms determined with
respect
to  the second fundamental form $k$ in order to have the well-defined
variational
problem. If $n^\mu$ is an outward  vector normal to
the boundary $\partial M$, then $k=\nabla_\mu n^\mu$.
The function $\psi$ is the solution
of the equation
\begin{equation}
\Box \psi= R
\label{1psi}
\end{equation}
where $\Box=\nabla_\mu\nabla^\mu$.

The action $I_2$ is  the Polyakov-Liouville term\footnotemark\footnotetext{The
reasons for writing
the Polyakov-Liouville term in the form (\ref{I2}) are analyzed in \cite{FIS}.}
incorporating both
the Hawking radiation of the scalar matter $N$-multiplet ($\kappa={N \over
24}$)
and its back-reaction on the black hole geometry. The local term $I_1$
(\ref{I1}) is added \cite{9d} to preserve, on the quantum level, some symmetry
of the classical action
(\ref{I0}). We are working in the semiclassical approximation when only the
matter
fields surrounding the black hole are  quantized while the metric of
two-dimensional space-time is still classical. Then, the minimum of the
effective
action
\begin{equation}
I=I_0+I_1+I_2
\label{I}
\end{equation}
under appropriately defined
 boundary condition gives us the quantum-corrected black hole configuration.

Varying (\ref{I}) with respect to metric we get the equation
($T_{\mu\nu}=2{\delta I \over
\delta g^{\mu\nu}}$):
\begin{equation}
T_{\mu\nu} \equiv
T_{\mu\nu}^{(0)}+
T_{\mu\nu}^{(1)}+
T_{\mu\nu}^{(2)}=0,
\label{T}
\end{equation}
where
\begin{equation}
T^{(0)}_{\mu\nu}={1 \over \pi}e^{-2 \phi} \left( 2 \nabla_\mu \nabla_\nu
\phi-2g_{\mu\nu} (\Box \phi-(\nabla \phi)^2+\lambda^2) \right)
\label{T0}
\end{equation}
\begin{equation}
T_{\mu\nu}^{(1)}=-{\kappa \over \pi} \left( g_{\mu\nu} \Box \phi -\nabla_\mu
\nabla_\nu \phi
\right)
\label{T1}
\end{equation}
\begin{equation}
T^{(2)}_{\mu\nu} =- { \kappa \over 2 \pi} \left( \partial_\mu \psi \partial_\nu
\psi
-2\nabla_\mu \nabla_\nu \psi -g_{\mu\nu} (-2R+{1 \over 2}(\nabla \psi)^2)
\right),
\label{T2}
\end{equation}
Variation of (\ref{I}) with respect to $\phi$ gives the dilaton field equation
\begin{equation}
2e^{-2\phi}(R+4 \Box \phi-4(\nabla \phi)^2+4\lambda^2)=-\kappa R
\label{dilaton}
\end{equation}
Taking trace of the energy-momentum tensor (\ref{T}) $T_{\mu\nu}g^{\mu\nu}=0$,
we get
\begin{equation}
2e^{-2\phi}(\Box \phi -2(\nabla \phi)^2+2\lambda^2)=-\kappa (R+\Box \phi)
\label{trace}
\end{equation}
Comparing (\ref{dilaton}) and (\ref{trace}) we come to the equation
\begin{equation}
(R+2\Box \phi)(\kappa-2e^{-2\phi})=0
\label{master}
\end{equation}

Remarkably, we now have only two possibilities.

The solution of the first type is characterized by the constant value of the
dilaton
\begin{equation}
2\phi=-\ln {\kappa \over 2} =  const
\label{phiconst}
\end{equation}
and the constant curvature
\begin{equation}
R=-2\lambda^2
\label{Rconst}
\end{equation}
Then equation (\ref{T}) reduces  to the theory of the induced 2D gravity with
the cosmological term (see  \cite{Polyakov}, \cite{Reuter}).
This de Sitter space-time  solution is absent in the classical theory
described by the action (\ref{I0}). Nevertheless, the value of the curvature
(\ref{Rconst}) is "classical" since it does not depend on  $\kappa$
characterizing the quantum effects ($\kappa$ is proportional to the
Planck constant $\hbar$).
This constant curvature solution was missed out in the previous consideration
of the model. Note that this solution lies completely in the quantum
mechanical strong coupling region. Therefore, one could assume that
it is an artifact of one loop and it is absent in the full quantum theory.
However, one can show that the
de Sitter space ($R=const$, $\phi=const$) is still the solution of three-loop
$\beta$-function equation in the $D=2$ $\sigma$-model \cite{AAT} that
 can be treated as $2D$ quantum gravity \cite{RT}, \cite{Banks-Lykken}.

\bigskip

{\large\bf B. Choice of boundary conditions}

The second possibility following from eq.(\ref{master}) consists in
that
the dilaton $\phi$ is a nonconstant function on the two-dimensional manifold.
Then, we have from (\ref{master}) the key relation\footnotemark\footnotetext{
This relation is present
in the classical theory described by the action $I_0$ (\ref{I0}). The one-loop
term
$I_1$ (\ref{I1}) is added in order to preserve this relation on the quantum
level.}
\begin{equation}
R=-2\Box \phi
\label{key}
\end{equation}
allowing one to integrate
exactly  all the field equations (\ref{T})-(\ref{dilaton}). Eq.(\ref{key})
means
that the function $\psi$ reads
\begin{equation}
\psi=-2\phi+w
\label{psi}
\end{equation}
where $w$ is the solution of the homogeneous equation, $\Box w=0$. The nonlocal
nature of
the action $I_2$ (\ref{I2}) is reflected in the dependence on such an arbitrary
function
$w$. The concrete choice of $w$ is provided by appropriate {\it  boundary
conditions}
corresponding to the chosen quantum state of the whole system.
One natural choice is to put $\partial_\mu\psi=0$ in the asymptotically flat
region. This means
that asymptotically $T^{(2)}_{\mu\nu}=0$ and, hence, no Hawking radiation is
present
in the flat Minkowskian
space-time. This boundary condition is reasonable in the situation when
formation
of a hole from flat space-time due to the incoming matter is considered
\cite{9d}.

Instead, here we assume the hole with non-zero mass to be already formed and
to be in the equilibrium with the environment of the fluctuating quantum fields
behaving like thermal gas at infinity.
The geometry of such eternal black hole is deformed by the back-reaction
 effects of this environment.
Our choice of the boundary conditions is regulated by the following two
requirements:
{\it 1)} There is no singularity of $\psi$ ( and $T^{(2)}_{\mu\nu}$ )
at the horizon for the Hawking temperature $T=T_H$.
{\it 2)} In the asymptotically flat region, the back-reaction is negligible and
we have  the semi-classical picture: flat space-time filled by the thermal
Hawking radiation with energy density $T^{0(2)}_{0}=T^0_{0,th}={\pi\over
6}NT^2$
and temperature $T$.

As has been shown in ref.\cite{SS}, the solution of the equation $\Box \psi =R$
for metric written in the Schwarzschild gauge
\begin{equation}
ds^2=-g(x)dt^2+{1 \over g(x)}dx^2
\label{metric}
\end{equation}
takes the form
\begin{equation}
\psi=- \ln g -b\int_{x}^{}{dx \over g(x)}
\label{psi0}
\end{equation}
and the renormalized energy density of the Hawking radiation is as follows:
\begin{equation}
T^{0(2)}_{0}={\kappa \over 2\pi} \left( 2g''_x -{1 \over 2g}(g'_x-b^2) \right)
\label{T00}
\end{equation}
The choice of the constant $b$ in (\ref{psi0}) and (\ref{T00}) specifies the
quantum
state of the system. It has been proposed in \cite{SS} that for the system
being at the temperature
$T=(2\pi \beta)^{-1}$ the natural choice is $b={2 \over \beta}$. Then, both
$\psi$
(\ref{psi0}) and $T^{0(2)}_{0}$ (\ref{T00}) occur to be regular at the horizon
($g(x_h)=0$) for the Hawking inverse temperature $\beta=\beta_H\equiv
2(g'(x_h))^{-1}$ and,
asymptotically, (\ref{T00}) gives the energy density of the thermal bath with
temperature $T$: $T^{0~(2)}_{0}\rightarrow T^0_{0,th}={\pi \over 6}N T^2$.
Thus, we have for the system lying in the box with size $L$ under  temperature
$T=(2\pi\beta)^{-1}$:
\begin{equation}
\psi=- \ln g -{2 \over \beta} \int_{x}^{L}{dx \over g(x)} +\ln g(L) +c
\label{psi'}
\end{equation}
where $c$ is constant.

\bigskip

The form (\ref{psi'}) for the function $\psi$ is general and valid for an
arbitrary 2D theory
describing eternal black hole.
Being interested in the concrete model (\ref{I}), it is instructive to give
 (\ref{psi'})
the semiclassical consideration  on the
classical black hole configuration \cite{5d} which is minimum of the action
$I_0$
(\ref{I0}). It takes  the form (\ref{metric}) with
\begin{equation}
g(\phi)=1-a e^{2 \phi}, ~~~\phi=-\lambda x
\label{classical}
\end{equation}
Then,  we obtain that at the Hawking temperature, $\beta=\beta_H\equiv {2 \over
g'(x_h)}$,
 the  function $\psi$  (\ref{psi'}) considered on this  classical
background takes the form (\ref{psi}) with $w=const$:
\begin{equation}
\psi (x)=-2\phi (x)+2\phi (L) +c,
\label{condition}
\end{equation}
where $\phi (L)$ is the value on the boundary $x=L$. Equivalently,
(\ref{condition})
means the condition $\partial_x \psi|_{x=L}=2\lambda$.

The contribution to the entropy due to the Hawking radiation is determined by
the value of function $\psi$ on the horizon \cite{12d}, \cite{FS}:
 $S=-2\kappa \psi (x_h)$. We obtain for
(\ref{condition}) that

$$
S=4\kappa (\phi (x_h)-\phi (L)-{c \over 2})
$$
Then, inserting  (\ref{classical}) we see that this quantity
\begin{equation}
S={\pi \over 3}N LT -2\kappa \ln a -2\kappa c
\label{ent}
\end{equation}
reproduces the entropy of the thermal bath $S_{th}=-4\kappa\phi (L)={\pi \over
3}NLT$ in the box
with size $L$ and temperature $T={\lambda \over 2\pi}$. The second term in the
r.h.s.
of (\ref{ent}), $4\kappa\phi (x_h)$ can be interpreted as an addition
(correction) to the entropy
of the black hole itself. Thus,  $\psi$ in the form (\ref{psi'}) with the
boundary
condition $\psi (x=L)=0$ (this fixes the indefinite constant in (\ref{psi'}),
$c=0$) includes automatically the effects of the
thermal bath of the asymptotically flat space-time.

We expect that the semiclassical consideration is correct in
 the asymptotically flat region. In particular, there we have (\ref{condition})
for
 $\psi$. Therefore, according
to our second requirement, we take
the gauge (\ref{condition}) or, equivalently, $w=const$ in (\ref{psi})
in the complete one-loop theory. The condition (\ref{condition}) will
guarantee, by the way,
the regularity of $\psi$ on the horizon.

In two dimensions the Hawking temperature and, correspondingly,
energy density of the Hawking radiation at infinity are independent of mass of
a hole. Therefore, our
requirements {\it 1)} and {\it 2)} concern only the nonzero mass  hole. For
flat space-time
(zero mass) there are no reasons for the radiation and, hence, we need
different
boundary condition, namely $\partial_\mu\psi \rightarrow 0$ ($T^{(2)}_{\mu\nu}
\rightarrow 0$) (or, equivalently, $w \rightarrow -2\lambda x$ in (\ref{psi})).

\bigskip

{\large\bf C. Exact integrability}

By taking into account (\ref{dilaton}) and (\ref{psi}), (\ref{condition}),
eq.(\ref{T})  is written in the form
\begin{equation}
\nabla_\mu \nabla_\nu F(\phi)={1 \over 2}g_{\mu\nu} \Box F(\phi)
\label{Feq}
\end{equation}
where $F$ is the function of the dilaton
\begin{equation}
F(\phi)\equiv \phi-{\kappa \over 4}e^{2\phi}
\end{equation}
An equation like (\ref{Feq}) normally appears in different two-dimensional
models of
gravity. It means that $\xi_\mu=\epsilon_{\mu}^{\ \nu} \partial_\nu F$
is the Killing vector ($\nabla_{( \mu} \xi_{\nu )}=0$).
This fact essentially simplifies  the integrating of field equations.
Indeed, we may use the variable $x={1 \over Q}F(\phi)$ as space-like coordinate
on the 2D space-time. Then, it follows from (\ref{Feq}) that metric takes the
Schwarzschild-like form (\ref{metric})
and it is static (independent of the time variable $t$).
The concrete form of the metric function $g(x)$ is found from  eq.(\ref{key})
which reads
\begin{equation}
\partial^2_x g=2\partial_x (g \partial_x \phi)
\label{key1}
\end{equation}
By integrating (\ref{key1}) it is more convenient to find $g$ as  a function of
the
dilaton $\phi$ under assumption that $\phi (x)$ is given by the equation
$Qx=F(\phi)$.
Then, we have from (\ref{key1}) that
\begin{equation}
2g=\partial_\phi g+{d \over Q} \partial_\phi F(\phi)
\label{difeq}
\end{equation}
where $d$ is  constant. The solution of (\ref{difeq}) is easily found:
\begin{equation}
g(\phi)=1 -a e^{2\phi} +\kappa \phi e^{2\phi}
\label{mfun}
\end{equation}
where we have put $d=2Q$ in order to have $g=1$ in the asymptotically flat
region
($\phi \rightarrow -\infty$); $a$ is the integrating constant.

Inserting now (\ref{metric}) and (\ref{mfun}) into (\ref{dilaton}), we get
$Q=-\lambda$
for the constant. Finally, the general solution of equations
(\ref{T})-(\ref{dilaton}) in the gauge (\ref{condition}) is the following
\begin{eqnarray}
&&ds^2=-g(x)dt^2+{1 \over g(x)}dx^2 ~~, \nn \\
&&- \lambda x=F(\phi) \equiv \phi-{\kappa \over 4}e^{2\phi}~~, \nn \\
&&g(\phi)=1-ae^{2\phi}+\kappa \phi e^{2\phi} ~~.
\l{solution}
\ea
Using the identity (\ref{difeq}) it is easy to check  that $\psi$ defined as
(\ref{psi'})
(at $\beta=\beta_H$) for the quantum-corrected background (\ref{solution})
indeed takes the
form (\ref{condition}). Thus, the whole integration procedure is
self-consistent.

As we could expect, the general solution (\ref{solution}) does not contain the
flat space.
Our boundary condition (\ref{condition}) assumes the presence  of the thermal
gas
with the nonzero energy density in the asymptotical region that necessary
curves
the space-time. As a  result, at infinity the positive thermal energy density,
$T^{0(2)}_0={\kappa\lambda^2 \over \pi}$,  is compensated by the negative
energy
density of the gravitational field\footnotemark\footnotetext{This is the energy
density of
the gravitational field described by the metric function $g_0=1+\kappa\phi
e^{2\phi}$ that is valid in the asymptotical region.},
 $T^{0(0)}_0=-{\kappa\lambda^2 \over \pi}$.

We see that in the limit $\kappa=0$ (\ref{solution}) coincides with
the classical black hole solution (\ref{classical}).
However, asymptotically ($\phi \rightarrow -\infty~, x\rightarrow +\infty $),
the last ("quantum") term in (\ref{solution}) dominates and the solution goes
not
to the classical one (\ref{classical}) but to that of (\ref{solution})
with $a=0$, $g\rightarrow g_0=1+\kappa\phi e^{2\phi}$. This
solution  is asymptotically flat and is a quantum deformation (with $\kappa$
being   the deformation parameter) of the classical linear dilaton vacuum.
It is an natural  reference configuration (instead of the flat space) with
respect to
which  the quantities (like energy) measured at infinity are defined.

\bigskip

\section{Global structure of the quantum corrected space-time}
\setcounter{equation}0

The dilaton field $\phi$ as a function of $x$ is two-valued.
The critical point $\phi_{cr}={1 \over 2} \ln {2 \over \kappa}$ defined
as $F'(\phi_{cr})=0$ separates  its two branches. So in the regions
$\phi < \phi_{cr}$ or $\phi > \phi_{cr}$, $\p (x)$ is one-valued.
We call these regions the $(+)$ and $(-)$ ones, respectively.
The derivative with respect to the variable $x$ is defined as $\partial_x
=-{\lambda
\over  F' (\p)} \partial_{\phi}$. Therefore, the point $\phi=\phi_{cr}$
is the place where the space-time singularity is present. Indeed, the scalar
curvature
for the metric (\ref{solution})
\begin{equation}
R\equiv -g''_x=-{4\lambda^2e^{2\phi} \over ({\kappa\over 2}e^{2\phi}-1)^3}
(a-\kappa-\kappa\phi +{\kappa^2\over 4} e^{2\phi})
\label{R}
\end{equation}
takes infinity at $\phi=\phi_{cr}$.

Thus, the singularity of the classical black hole ($\kappa=0$), located at
$\phi=+\infty$ is now shifted to the finite value of the dilaton, $\p =
\p_{cr}$.

Other important point characterizing (\ref{solution}) is that the flat
space-time
is not a solution to any parameter $a$. This is obviously due to the boundary
condition (\ref{condition}) and the fact that back-reaction of the Hawking
radiation
drastically changes the geometry of space-time.

The structure of space-time described
by the metric (\ref{solution}) essentially depends on the value of the
integrating constant
$a$ (in the next section we relate it with  the hole mass).

For $a<a_{cr}={\kappa \over 2}(1- \ln {\kappa \over 2})$ the metric function
$g(\p)$ is everywhere positive. So no horizon is present and (\ref{solution})
describes space-time with naked singularity.

At the critical value of $a=a_{cr}$ we have $g(\phi_{cr})=g'(\phi_{cr})=0$.
 Nevertheless, $g(x)$ has a simple zero
at $x_{cr}=-{1 \over \lambda} F(\p_{cr})$ since $g'_x (x_{cr})=2\lambda$.

For $a>a_{cr}$ the metrical function $g(\p)$ has two zeros, $\p_h$ and ${\tilde
\phi}_h$.
However, since $\p_h < \p_{cr} <{\tilde \phi}_{h} $, the second zero,
$\tilde{\phi}_h$,
is "beyond" the singularity.

Hence, for $a \geq a_{cr}$ in the region $\p <\p_{cr}$ the horizon appears.
For $a=a_{cr}$ it coincides with the singularity. In the classical case the
point $a=0$ separates
the solutions with and without horizons, the case $a=0$ corresponding to
everywhere
regular flat space-time (vacuum of the theory).
The quantum-corrected space  with $a=a_{cr}$ is also the
smoothest among other solutions (\ref{solution}). Indeed, for $a>a_{cr}$
the metric function $g(x)$ at $x=x_{cr}$ reads
\begin{equation}
g(x)=-{2 \over \kappa}(a-a_{cr})+2\lambda (1+{2 \over \kappa}(a-a_{cr}))
(x-x_{cr})-{4 \sqrt{\lambda} \over \kappa}(a-a_{cr})(x-x_{cr})^{1 \over 2}
\end{equation}
and both
$g'_x$ and $g''_x$ are singular at $\p=\p_{cr}$, the curvature going
to infinity as follows:
\begin{equation}
R \sim -{\lambda^2 (a_{cr}-a) \over \kappa (\p_{cr}-\p)^3}=
-{\lambda^{1 \over 2} (a_{cr}-a) \over \kappa (x_{cr}-x)^{3 \over 2}}
\l{R1}
\end{equation}
For $a=a_{cr}$ the metric
\begin{equation}
g(x)=2\lambda (x-x_{cr})+{8 \over 3} \lambda^{3 \over 2} (x-x_{cr})^{3 \over 2}
\end{equation}
and the first derivative $g'_x$ are regular at $\p=\p_{cr}$, while
the curvature (or second derivative $g''_x$)
\begin{equation}
R \sim -{\lambda^2 \over (\phi-\phi_{cr})}=-{\lambda^{3 \over 2} \over
 (x-x_{cr})^{1 \over 2}}
\l{R2}
\end{equation}
One can see that the singularity (\ref{R1}) is stronger than (\ref{R2}).
Generally, the metric (\ref{solution}) is  smoother in comparison
with the classical one. In the latter, the singularity is exponential
$R=-4\lambda^2ae^{2\phi}$ while in the quantum-corrected
case curvature grows by power law (\ref{R1}), (\ref{R2}).
Moreover, the classical singularity manifests itself already in the
singular behavior
of the metric function, $g_{cl} \sim -ae^{2\p}$, while the quantum-corrected
metric function (\ref{solution}) is regular at $\p=\p_{cr}$ and only
derivatives , $g'_x$ and $g''_x$ diverge. This circumstance allows us
to formally consider the regions $\p < \p_{cr}$ and $\p > \p_{cr}$ as different
sheets of the same space-time glued at $\p= \p_{cr}$. The coordinate
$\phi$ naturally parametrizes both sheets while $x$ is appropriate to
giving a coordinate only on one of them.
Both the sheets are asymptotically flat though the curvature reaches zero
differently:
\ba
&&R \sim -4\kappa \lambda^2 \p e^{2\p} ~~, ~\p \rightarrow -\infty ~~~(+)-sheet
\nn \\
&&R \sim -{8\lambda^2 \over \kappa} e^{-2\phi}~~, ~\p \rightarrow +\infty ~~~
(-)-sheet
\l{R3}
\ea
For $a>a_{cr}$ every sheet contains an event horizon: $\p_h$ on the $(+)$-sheet
and ${\tilde \phi}_h$ on the $(-)$-sheet. Remarkably, the derivative of the
metric $g'_x$
is the same for both the horizons and is equal to $g'_x(x_{h})=2\lambda$.

Of course, this picture looks formal since no any observer can penetrate
through the singularity at $\p=\p_{cr}$ and appear at the second sheet.
However,
this picture is mainly the result of  the back-reaction effects within the
one-loop
approximation. We assume that next loops taken into account will
make the singularity at $\phi=\phi_{cr}$ smoother preserving the general global
structure
to be the same.
So, having taken the full effective action, which is the result of all loops
contributions, the singularity could be expected to vanish completely.

 One can find some support to this idea in the study of the exact
(non-perturbative)
two-dimensional space-time \cite{VV} that emerges from  string theory
as an exact background of the string target space. The analysis of its global
structure
shows \cite{PT} the remarkable picture: two copies of the black hole space-time
having an event horizon but no singularity are glued together to form a
wormhole
bridging two asymptotically flat regions. This is in agreement with our present
consideration. It would be interesting to find the form of the corresponding
(non-perturbative) gravitational action of, possibly, dilaton type giving the
dynamics of the string target space geometry and possessing this kind of
solution.

\bigskip

\section{Thermodynamics, mass and entropy formulas}
\setcounter{equation}0

The quantum-corrected black hole solution (\ref{solution}) resulted
from the one-loop quantum effects. Generally, one could expect that these
effects
lead also to  modification of all characteristics of the hole (mass, entropy,
temperature) that  possesses quantum corrections. This could
change thermodynamical relations, say, entropy as a function of mass, etc.

In this section, we study this problem for the RST model and for the
quantum-corrected
black hole solution found. To seek the completeness we begin our analysis with
general remarks on  the formulation of the black hole thermodynamics
and brief description of the thermodynamics of the classical hole
(\ref{classical})
(see \cite{20}, \cite{21}, \cite{22}, \cite{SS}).

\bigskip

{\large\bf A. Formulation of the black hole thermodynamics}

Consider the system (gravity plus matter)
at arbitrary temperature $T=(2\pi\beta )^{-1}$.
The thermodynamics of the field system  usually has the Euclidean
formulation making the Wick rotation $t=\imath \tau$ and supposing for all
fields
to be periodical with respect to imaginary time $\tau$ with the period
$2\pi\beta =T^{-1}$, where $T$ is temperature of the system.

We define the state of the system  as any configuration $(\phi (x),
g_{\mu\nu}(x))$
satisfying some general conditions: {\it a)} $\phi (x)$ and $g_{\mu\nu}(x)$ are
real
fields on the Euclidean  manifold $(t=\imath \tau )$ with abelian isometry
along
the Killing vector $\partial_\tau$ with the period $2\pi\beta$; {\it b)}
There exists a subspace (horizon) which is a fixed point of the isometry;
{\it c)} metric $g_{\mu\nu}(x)$ is asymptotically flat.

The condition {\it a)} in two dimensions means that metric  can be written in
the
form
\begin{equation}
ds^2=g(x)d\tau^2+g^{-1}(x)dx^2,
\label{4.1}
\end{equation}
where $0 \leq \tau \leq 2\pi\beta$.

{}From condition {\it b)} it follows that the metric function has zero
$g(x_h)=0$
at some point $x=x_h$. This means that the system includes  black hole with the
horizon
at $x=x_h$.  According to {\it c)} the metric function goes to $g(x)\rightarrow
1$
if $x\rightarrow \infty$. It should be noted that any other constraint on the
state
$(\phi (x),g_{\mu\nu}(x))$ is not assumed. In particular, values on the horizon
$\phi (x_h),g'_x(x_h)$ are arbitrary. We only assume that the system includes
the
nonextremal hole, i.e. $g'(x_h) \neq 0$.
This is essential that if $\beta \neq \beta_H\equiv {2 \over g'(x_h)}$, then
(\ref{4.1})
describes space with conical singularity on the horizon. This singularity
manifests itself in the $\delta$-like contribution to the curvature so that the
complete
quantity reads \cite{FS}:
\begin{equation}
\bar{R}=2({1-\alpha \over \alpha})\delta (x-x_h)+R, ~~~\alpha={\beta \over
\beta_H}
\label{4.2}
\end{equation}
where $R=-g''$ is the regular part of the curvature.
Thus, our statistical ensemble contains both the regular and singular Euclidean
metrics\footnote{In this point our approach differs from that developed by York
with collaborators \cite{Y} in which only {\it regular } spacetimes of black
hole topology are suggested to form a statistical ensemble. Nevertheless, for
quantities calculable at the Hawking temperature the both approaches give the
same results.}.

With respect to the action $I[\beta,g,\phi ]$ one can define the free energy
$F={1 \over 2\pi\beta}I$ which is functional, $F=F[\beta,g_{\mu\nu}(x),\phi
(x)]$,
of the fixed inverse temperature $\beta^{-1}$ and of the state $(\phi (x),
g_{\mu\nu}(x))$.

Applying the thermodynamical formulas
\begin{equation}
S=(\beta \partial_\beta-1)I [\beta],~~~E={1 \over 2\pi} \partial_\beta I
[\beta]
\label{4.4}
\end{equation}
we  may calculate the energy and entropy for an arbitrary state $(\phi
(x),g_{\mu\nu}(x))$
at the fixed $\beta$. These quantities for a system being at the fixed
temperature
change until a system reaches a thermal equilibrium characterized by the
extremum of
the free energy $F=E-TS$ (or, equivalently,  of the action $I[\beta, \phi
(x),g_{\mu\nu} (x)]$),
$(\delta F)_\beta=0$. In this variational problem, as it follows from the
conditions
{\it a)-c)}, only behavior of fields $\phi , g$ at infinity or at the boundary
of the box ($x=L$) is fixed, $\delta\phi|_{x=L}=\delta g|_{x=L}=0$.
Remarkably, such a equilibrium configuration satisfies
the 2-nd law of thermodynamics
\begin{equation}
\delta E=T \delta S
\label{4.0}
\end{equation}
for small variations around the equilibrium state.

This extreme configuration satisfies the field equations obtained from the
action $I$,
$\delta_g I=\delta_\phi I=0$, and for  all known cases the extremum is reached
on
the regular manifold, i.e., the corresponding Hawking temperature coincides
with
the fixed temperature of the system $\beta=\beta_H$. Thus, the state of the
system
evolves until its Hawking temperature becomes equal to the  temperature of the
system fixed  from the beginning.

The entropy (\ref{4.4}) taken at  $\beta=\beta_H$ and satisfying (\ref{4.0})
is the Bekenstein-Hawking entropy which is determined by total response of the
free energy $F$ of the system being in the thermal equilibrium on variation of
temperature\footnotemark\footnotetext{This follows from the condition $(\delta
F)_\beta=0$ defining
the equilibrium state.}.

\bigskip

{\large\bf B. Thermodynamics of classical black hole}

Apply now these prescriptions
to the classical black hole
described by the action $I_0$ (\ref{I0}) (after Wick rotation
it changes the overall sign) which on an arbitrary metric with conical
singularity
takes the form
\begin{equation}
I_{0}[\beta]=-2e^{-2\phi_h}(1-{\beta \over \beta_H})-
{1 \over 2\pi} \int_{M}^{}d^2 x \sqrt{-g} e^{-2\phi}
[R+4(\nabla \phi)^2+4\lambda^2]-{1 \over \pi}\int_{\partial M}^{}e^{-2\phi}kds
\label{4.3}
\end{equation}
It is a functional of the inverse temperature $\beta$.
Consider the configuration which
minimizes the action functional  (\ref{4.3}) under $\beta$ fixed. This would be
an equilibrium configuration for the given temperature. It should be noted that
only
a large distance behavior  of the metric $g(L)\rightarrow 1,~  L\rightarrow
\infty$
is assumed to be fixed in this variational problem. The functions $g(x),g'(x),
\phi(x)$
and values on the horizon $\phi (x_h),~~g'(x_h)={2 \over \beta_H}$ are supposed
to
be variable. As a result, the variation with respect to $\delta \phi (x_h)$
gives
the constraint
\begin{equation}
{2 \over g'(x_h)}\equiv \beta_H=\beta
\label{4.5}
\end{equation}
It means that the equilibrium configuration is a regular manifold without
conical
singularities.

The variation of (\ref{4.3}) with respect to $\delta g'(x_h)$ vanishes
automatically
due to mutual cancellation of variation  of $\beta_H$ in the first term defined
on the horizon
$x=x_h$ and that of coming from the second term in the r.h.s. of (\ref{4.3}).
Other (volume) variations  give the classical dilaton (eq.(\ref{dilaton}) with
$\kappa=0$) and gravitational $T^{(0)}_{\mu\nu}=0$ equations. The
solution is given by (\ref{classical}).

For the equilibrium state we have the temperature $T={1 \over
2\pi\beta_H}={\lambda \over 2\pi}$
and the Bekenstein-Hawking entropy of the classical
hole\footnotemark\footnotetext{Translating the
two-dimensional physics to the $4D$ language it is useful to have in mind
the analogy between the dilaton field $\phi$ and the radius $r^2$ in the $4D$
spherically symmetric case, $r^2 \sim e^{-2\phi}$ (see the Section 5 for
details).
Then, (\ref{4.6}) is written as $S\sim r^2_h$ that is similar to the known
four-dimensional law relating entropy of a hole with the area of the horizon
$A_h=\pi r^2_h$.}
\begin{equation}
S_{BH}=2e^{-2\phi_h}=2a
\label{4.6}
\end{equation}
On the constraint $T^{(0)}_{00}=0$  the energy functional (\ref{4.4}) reduces
to the boundary term
\begin{eqnarray}
&&E={1 \over 2\pi\beta}\int_{\partial M}^{}{2\over \pi}e^{-2\phi}  n^\mu
\partial_\mu \phi ds \nonumber \\
&&={2 \over \pi} (e^{-2\phi}g(x)\phi'_x)_{x=L}
\label{4.7}
\end {eqnarray}
where $ds=g^{1/2} dx$ is measure induced on the boundary.
This quantity is divergent in the limit $L\rightarrow \infty$ for the classical
solution
(\ref{solution}) and, in particular, for the flat space ($g=1, \phi=-\lambda
x$).
To regularize the quantity (\ref{4.7}), one must subtract the flat space
contribution.
To this aim, we  add to the action $I_0$ the additional boundary term
\begin{equation}
I_0 \rightarrow I_0 -{2 \over \pi}\int_{\partial M}^{}e^{-2\phi}(n^\mu_0
\partial_\mu
\phi )ds
\label{4.8}
\end{equation}
where the normal vector $n^\mu_0$ is defined with respect to the flat metric
$g=1$. Then, expression for the energy (\ref{4.7}) gets modified:
\begin{equation}
E={2 \over \pi} \left(e^{-2\phi} g^{1/2}(g^{1/2}-1)\phi' (x) \right)_{x=L}
\label{4.9}
\end{equation}
Substituting solution (\ref{classical}) into (\ref{4.9}) we obtain in the limit
$L\rightarrow \infty$
\begin{equation}
E={a\lambda \over \pi}
\label{4.10}
\end{equation}
This is the well-known result (\cite{20}, \cite{21}, \cite{22}) for the mass
of the dilaton black hole.

\bigskip

{\large\bf C. Thermodynamics of the quantum-corrected black hole}

The same approach can be applied to the formulation of  thermodynamics of the
quantum-corrected hole described by the action (\ref{I}). To get
the terms $I_0$ and $I_1$ on the manifold with conical singularity we can again
use formula (\ref{4.2}) for the complete curvature. One must be more careful,
however, with the Polyakov-Liouville term $I_2$. It is obtained by
integrating the conformal anomaly. It should be noted that the anomaly
becomes modified due to the contribution from the conical
singularity\footnotemark\footnotetext{ The conformal
anomaly is determined by the heat kernel coefficient $a_2$, $\delta_\sigma W=
\int_{M}^{}\delta \sigma a_2(x)$, which in the presence of the conical
singularity has
a $\delta$-like contribution from the tip of the cone \cite{F}.}
that really modifies  the action $I_2$. However, comparison with the exact
results shows
that up to $(1-\alpha)^2$ terms this coincides with naive applying of
(\ref{4.2})
to  (\ref{I2}).

Taking into account that $\psi=-2\phi +C$, where $C$ is constant, we finally
come to
the quantum action $I$ for the metric (\ref{4.1}) for  arbitrary $\beta$:
\begin{eqnarray}
&&I=-{1 \over 2\pi}\int_{M}^{}\left(
(e^{-2\phi}+\kappa\phi)R+2(2e^{-2\phi}-\kappa)
(\nabla\phi)^2+4\lambda^2e^{-2\phi}\right) \nonumber \\
&&-{1 \over \pi}\int_{\partial M}^{}(e^{-2\phi}+\kappa\phi)k-2(1-{\beta\over
\beta_H})
(e^{-2\phi_h}+\kappa\phi_h)+2\kappa C \chi (M),
\label{4.11}
\end{eqnarray}
where $\phi_h=\phi (x_h)$ is the value on the horizon, $\beta_H={2 \over
g'(x_h)}$
and

$$
\chi (M)={1 \over 4\pi}(\int_{M}^{}R+2\int_{\partial M}^{}k)+(1-{\beta\over
\beta_H})
$$
is the Euler number of the manifold $M$ for arbitrary $\beta$ \cite{SS},
\cite{FS}.  One can easily see
that $\chi (M)=1$.

Fixing $\beta$ and varying  all the variables $g,g',\phi$ we again obtain that
equation
$\delta_{\phi_h}I=0$ gives the condition (\ref{4.5})
while $\delta_{\phi (x)}I=\delta_{g(x)}I=0$ are the field equations
(\ref{T})-(\ref{dilaton}).
So the quantum-corrected equilibrium state is again a regular configuration
at the Hawking temperature, $\beta=\beta_H$. As it can be seen from
(\ref{solution}),
on the horizon we have $g'(x_h)=2\lambda$. Hence, the temperature of the
quantum-corrected
hole is the same as for the classical hole,
$\beta_H=\lambda^{-1},~~T_H={\lambda \over 2\pi}$.

{}From  (\ref{4.11}) we obtain the expression for the entropy
\begin{equation}
S=2(e^{-2\phi_h}+\kappa\phi_h)-2\kappa C
\label{4.12}
\end{equation}
where $C=-2\phi (L)+c$ ( see (\ref{condition})). The first term in the r.h.s.
of (\ref{4.12})
is defined on the horizon and can naturally be interpreted as the
Bekenstein-Hawking
entropy of the hole itself\footnotemark\footnotetext{Taking the analogy with
the $4D$ spherically symmetric case,
this formula means a modification of the  {\it entropy-area} relation by a
logarithmic correction $S\sim A_h-\kappa\ln A_h$.}:
\begin{equation}
S_{BH}=2(e^{-2\phi_h}+\kappa\phi_h)
\label{4.13}
\end{equation}
This result up to an additive constant coincides with that of
previously obtained in \cite{11d}-\cite{13d}.
On the other hand, the second term in (\ref{4.12})  coincides (for $c=0$)
with the entropy of the thermal bath with temperature $T_H={2\pi\over
\lambda}^{-1}$
filling the space outside the horizon
\begin{equation}
S_{th}=-4\kappa\phi (L)={\lambda N L\over 6}
\label{4.14}
\end{equation}
Thus,  the method of the conical singularity being applied to the quantum
effective
action gives us both the Bekenstein-Hawking entropy of the quantum-corrected
hole
and the entropy of thermal gas surrounding the hole. This naturally happens
when we
put the appropriate boundary condition $\psi (L)=0$ for the function $\psi (x)$
playing an important role in the two-dimensional quantum gravitational physics.
In principle, we may subtract the entropy of the gas $S_{th}$ in the complete
entropy
putting $c=2\phi (L)~~(C=0)$ in (\ref{condition}). Generally, different
constants
$c$ correspond
to different choices of the  reference point for computation of the
system's entropy.

Substituting the solution (\ref{solution}) into (\ref{4.13}) we obtain that the
entropy
of the quantum-corrected hole coincides with the classical one
\begin{equation}
S_{BH}=2a
\label{4.15}
\end{equation}
Measuring the entropy with respect to that of  vacuum  defined as solution
(\ref{solution}) for $a=a_{cr}$ we obtain
\begin{equation}
S_{BH}=2(e^{-2\phi_h}+\kappa\phi_h )-2 (e^{-2\phi_{cr}}+\kappa\phi_{cr} )=
2(a-a_{cr})
\end{equation}
that exactly coincides with the expression derived in \cite{11d}-\cite{13d}.

Using the constraint $T_{00}=0$ we obtain  (after short calculations) the
expression
for the energy
\begin{eqnarray}
&&E={1 \over 2\pi\beta}{1 \over 2\pi}\int_{\partial M}^{}(2e^{-2\phi}- \kappa)
n^\mu\partial_\mu \phi ds \nonumber \\
&&={1 \over \pi}[(2e^{-2\phi}-\kappa)g\phi' ]_{x=L}
\label{4.16}
\end{eqnarray}
Considering (\ref{4.16}) on  (\ref{solution}) we see that $E$ is divergent in
the
limit $L\rightarrow \infty$. Subtracting from the action the same boundary term
(\ref{4.8}) as in the classical case  we obtain that the energy
\begin{equation}
E={1 \over \pi}[(2e^{-2\phi}-\kappa )g\phi'-2e^{-2\phi}g^{1/2}\phi' ]_{x=L}
\label{4.17}
\end{equation}
is still divergent
\begin{eqnarray}
&&E={\lambda a \over \pi}-{\kappa \lambda \over \pi} \phi (L) +{\lambda\kappa
\over \pi}
\nonumber \\
&&={\lambda a \over \pi}+{\lambda^2 N L \over 24 \pi}+{\lambda \kappa \over
\pi}
\label{4.18}
\end{eqnarray}
Up to the last, irrelevant, term   eq.(\ref{4.18}) can be interpreted as a sum
of mass of the hole
itself,
\begin{equation}
M={\lambda a \over \pi}
\label{4.19}
\end{equation}
and of the (divergent) energy of the thermal gas surrounding the hole
\begin{equation}
E_{th}={\lambda^2 N L \over 24\pi}
\label{4.20}
\end{equation}
We  may exclude this contribution of thermal gas and
obtain the finite energy if instead of (\ref{4.8}) we subtract in the action
the quantum-corrected boundary term:
\begin{equation}
I\rightarrow I-{1\over \pi}\int_{\partial M}^{}(2e^{-2\phi}-\kappa\phi )n^\mu_0
\partial_\mu \phi,
\label{4.21}
\end{equation}
where the normal vector $n^\mu_0$  is defined with respect to the
metric $g_0$, the solution (\ref{solution}) corresponding to $a=0$, which
replaces
the flat space at large distances in the quantum case.
Then the energy reads
\begin{equation}
E={1 \over \pi}[(2e^{-2\phi}-\kappa)g^{1/2}(g^{1/2}-g^{1/2}_0)\phi'_x]_{x=L}
\label{4.22}
\end{equation}
where $g_0=1+\kappa \phi e^{2\phi}$. This quantity is finite and equal to
(\ref{4.19}),
$E=M$.

We may measure the mass of the hole with respect to vacuum (solution
(\ref{solution})
for $a=a_{cr}$). Then, in the boundary term (\ref{4.21}) the normal vector
$n^\mu_0$ must be defined with respect to the vacuum metric. The expression for
energy takes the form (\ref{4.22}) where now $g_0=g_{vac}\equiv
1-a_{cr}e^{2\phi}+\kappa\phi
e^{2\phi}$. The resulting energy

$$
E={\lambda \over \pi}(a-a_{cr})
$$
differs form (\ref{4.19}) by a constant and vanishes for the vacuum
configuration
($a=a_{cr}$).

\bigskip

There is an alternative derivation of the mass advocated in
\cite{22}.  Let us assume that all field equations except the gravitational one
(\ref{T}) are satisfied. The coordinate invariance of the action (\ref{I})
implies that
\begin{equation}
\nabla_\mu T^{\mu\nu}=0
\l{law}
\end{equation}
where $T_{\mu\nu}$ is given by (\ref{T})-(\ref{T2}). Let the gravitational
field be
 static and allow a timelike Killing vector $\xi_\mu$. Then, one has
\begin{equation}
\nabla_\mu (T^{\mu\nu} \xi_\nu )=0
\l{law1}
\end{equation}
In two dimensions (\ref{law1}) implies that there exists such a scalar function
$M$ that
\begin{equation}
\nabla_\alpha M=-\epsilon_{\alpha\mu}T^{\mu\nu}\xi_\nu~~~.
\l{M}
\end{equation}
In our case $\xi_\mu =-{1 \over \lambda} \epsilon_{\mu}^{\ \nu} \partial_\nu
F(\p)$ and in the gauge (\ref{psi})-(\ref{condition}) after simple calculations
we get
\begin{equation}
M={\lambda \over \pi}(\kappa \p+ e^{-2\p})-{1 \over \pi\lambda} e^{-2\p}
(\nabla F)^2
\l{M1}
\end{equation}
When the gravitational field equations $T_{\mu\nu}=0$ (\ref{T}) are satisfied,
eq.(\ref{M1}) implies that $M=const$. So (\ref{M1}) gives the first integral
of the gravitational field equations. Indeed, for the configuration
(\ref{solution})
we obtain the result (\ref{4.19}), $M={\lambda a \over \pi}$.
It is worth noting that (\ref{M1}) allows one to write the mass formula,
relating $M$ with values at the horizon:
\begin{equation}
M={\lambda \over \pi}(\kappa \p_h +e^{-2\p_h})
\l{M4}
\end{equation}

\bigskip

{\large\bf D. Comparison with perturbative calculations}

The most important conclusion  from the above consideration is that in the
exact
one-loop semiclassical theory the thermodynamical characteristics
(the Hawking temperature, mass and entropy) of the quantum-corrected black hole
do
not possess any quantum corrections. So all the characteristics  coincide with
the classical ones.

In the previous  consideration within the perturbative approach with respect to
$\kappa$ (or, equivalently, the Planck constant) the logarithmic, $\ln M$,
correction
to the entropy has been observed \cite{11d}, \cite{SS}. Therefore, it is
worth comparing these perturbative results with the present exact calculation.

On the one-loop level the classical expression for the Bekenstein-Hawking
entropy
(\ref{4.6}) as a function of  dates  on the horizon is modified by the term in
(\ref{4.13})
proportional to $\kappa$. This additional term can be treated as quantum
correction, $S_q=2\kappa \phi_h$. Expanding the value on the horizon with
respect to
$\kappa$, $\phi_h=\phi_h^{cl}+\kappa\phi_h^{q}$, we obtain
\begin{equation}
S_q=2\kappa\phi_h^{cl}=-\kappa \ln {\pi M \over \lambda}
\label{4.27}
\end{equation}
This is the result obtained in \cite{11d}, \cite{SS} and interpreted as a
quantum correction
to the classical entropy (\ref{4.6}). The correction (\ref{4.27})
is essentially due to modification of the entropy formula in the one-loop
theory.
However, in the consistent approach we must also expand the first ("classical")
term\footnotemark\footnotetext{I thank V.P.Frolov for this important remark.}
$S_{cl}=2e^{-2\phi_h}$ in (\ref{4.13}) with respect to $\kappa$
\begin{equation}
S_{cl}=2e^{-2\phi_h^{cl}}(1-2\kappa\phi^q_h)
\label{4.28}
\end{equation}
{}From (\ref{solution}) we obtain that $2\phi^q_h=\phi^{cl}_he^{2\phi_h^{cl}}$.
Then, (\ref{4.28}) reads
\begin{equation}
S_{cl}=2e^{-2\phi_h^{cl}}-2\kappa\phi_h^{cl}
\label{4.29}
\end{equation}
This correction is due to the deformation of the geometry of the black hole and
of the
horizon "location", $\phi_h$, in the one-loop theory.
We see that both the one-loop modifications: of the entropy formula and black
hole geometry, are mutually compensated and the complete entropy, which is the
sum of (\ref{4.27}) and (\ref{4.28}), remains
uncorrected\footnotemark\footnotetext{This,
however, does not exclude that entropy as a function of $\phi_h$ gets
modified.
Taking analogy $r^2\sim e^{-2\phi}$, this means that entropy as
a function of the horizon area is really modified by quantum corrections.
}.

There are arguments  similar to that leading to (\ref{4.27}) concerning
 quantum corrections to entropy of the four-dimensional black hole \cite{F1}.
However, our present consideration shows that
one-loop calculations on the fixed {\it classical} background must be
accompanied
by an analysis of
the changing of the black hole geometry due to the back-reaction effects.
The latter can be important  to make the final conclusion about quantum
corrections
to the black hole thermodynamics.

\bigskip

\section{Relevance to four dimensions}
\setcounter{equation}0

In four dimensions we face much more difficult problem.
At first, we do not have  complete knowledge about the one-loop
effective action (more exactly, about its finite nonlocal part).
Therefore,  modifications of the mass and entropy formulas are not
exactly known. On the other hand, attempts to find  quantum-corrected
solutions minimizing the effective action look hopeless. However,
many things are simplified when space-time symmetries are present.
For example, the structure of the renormalized energy-momentum tensor
for the static spherically symmetric background has been studied in
 more details  \cite{Page}. This allows one to find
a quantum-corrected black hole configuration in some approximation
\cite{Sanchez-Lousto}.
In this Section we illustrate the consideration of the previous section in a
somewhat
different approach developed in \cite{15d} and allowing the reduction of the 4D
problem to the two-dimensional one.

Indeed, the gravitational Einstein-Hilbert action
\begin{equation}
I_{gr}={1 \over 16\pi G}\int_{}^{}d^4 x \sqrt{-g} R^{(4)}
\label{EH}
\end{equation}
being considered on the four-dimensional spherically symmetric metric
of general type:
\begin{equation}
ds^2=\sum_{\alpha , \beta=0}^{1}g_{\alpha\beta}^{(2)} dz^\alpha dz^\beta
+r^2(z) (d\theta^2 +\sin^2 \theta d \varphi^2 ),
\label{sph}
\end{equation}
is reduced to the effective two-dimensional theory
\begin{equation}
I_{gr}={1 \over 4G} \int_{}^{}dz^2 \sqrt{-g^{(2)}} \left( r^2 R^{(2)}+
2( \nabla r)^2 +2 U(r) \right)
\label{EH1}
\end{equation}
of the dilaton gravity type. The ${r^2 \over G} \equiv e^{-2\phi}$ plays the
role
of the dilaton field. The "dilaton" potential $U(r)$ is constant $U(r)=1$
in the classics. Quantizing  only the
spherically symmetric excitations in the original theory (\ref{EH})  we come
to the quantum theory of dilaton gravity
(\ref{EH1}). The potential $U(r)$ changes its form due to quantum corrections
that was found by solving the corresponding renorm-group equation
(for details see  \cite{15d}). The ultraviolet divergences have been shown to
be
absorbed in the redefinition  of the gravitational coupling $G$. Though in
\cite{15d} we considered the more general case, we restrict ourselves here,
just for illustration, by simple case when one neglects the possible anomalous
terms in the quantum version of action (\ref{EH1}). This approximation
is good for enough large mass of hole, $M > 10M_{pl}$. The corrected dilaton
potential then reads
\begin{equation}
U(r)={r \over (r^2-l^2_{pl})^{1 \over 2}}
\end{equation}
where $l^2_{pl}=8G_{ren}$ is distance of the Planck order.
Then the quantum corrected metric takes the form
\begin{equation}
ds^2=-g(r)dt^2+{1 \over g(r)}dr^2+r^2 d\Omega^2
\label{m1}
\end{equation}
where
\begin{equation}
g(r)=-{2M \over r}+{\sqrt{r^2-l^2_{pl}} \over r}
\label{g}
\end{equation}
For $r>>l_{pl}$ the classical Schwarzschild black hole solution restores.

Remarkably, the space-time described by the metric (\ref{m1}), (\ref{g})
is quite similar to the two-dimensional space-time (\ref{solution})
considered in the previous Sections.
Indeed, (\ref{m1}), (\ref{g}) can be written in the form similar to
(\ref{solution}):
\begin{eqnarray}
&&ds^2=-g(r)dt^2+g^{-1}(r)dr^2+r^2(\phi)d\Omega^2, \nonumber \\
&&{r \over l_{pl}}=F(\phi)\equiv \cosh \phi, \nonumber \\
&&g(\phi)= th \phi -2M (\cosh \phi )^{-1}
\label{gp}
\end{eqnarray}
The classical singularity at $r=0$
($\phi=+\infty$) is now shifted to the finite distance $r=l_{pl}$
($\phi=\phi_{cr}$).  It has been proposed in \cite{15d}, that the metric
(\ref{m1}), (\ref{g}) is formally  extended  behind the singularity $r=l_{pl}$
to the second asymptotically flat sheet ($\phi \rightarrow +\infty$)
and the singularity
becomes smoother due to contributions of ghosts and matter fields.
The general structure of the full
space-time is similar to that we discussed in Section 3.
The essential difference from (\ref{solution}) is that there is no extra
horizon on the second sheet.

At large $r>>l_{pl}$ the space-time is no more Ricci flat and the curvature
falls as follows:
\begin{equation}
R^{(4)}=-{2 \over r^2}({l_{pl} \over r})^4
\end{equation}
The second term in the r.h.s. of (\ref{g}) at large $r$ falls as $1+O({1 \over
r^2})$.
Therefore, the mass $M$ of the hole does not possess any corrections.
Nevertheless, the horizon defined by $g(r_h)=0$ is now shifted
\begin{equation}
r_h=\sqrt{(2M)^2+l^2_{pl}}
\l{horizon}
\end{equation}
compared with the classical one $r_{h,cl}=2M$. However, the Hawking temperature
$T_H={g'(r_h) \over 4\pi}$ remains unchanged
\begin{equation}
T_H={1 \over 8\pi M}
\l{TH}
\end{equation}
The entropy derived from the action (\ref{EH1}) reads
\begin{equation}
S={A_h \over 4G}
\l{SQ}
\end{equation}
where $A_h=4\pi r^2_h$ is the area of the quantum corrected horizon
(\ref{horizon})
and $G$ is the renormalized gravitational coupling.
One can see that $S$ (\ref{SQ})being expressed via the corrected quantities
takes the classical Bekenstein-Hawking form. In terms of the classical horizon
area, however, one observes the constant correction
to the classical law
\begin{equation}
S={A_{h,cl} \over 4G}+\eta
\l{SQ1}
\end{equation}
The concrete value of the constant $\eta=8\pi$ is irrelevant.
Expressions (\ref{TH})-(\ref{SQ1}) illustrate the idea that thermodynamical
characteristics of a hole ($M, S, T_H$ etc.) being expressed in terms of
the quantum-corrected quantities may take the classical form. As we have seen,
this is realized
for the black hole in the two-dimensional RST model. In this Section we found
also
the partial support to this idea for the $4D$ Schwarzschild black hole.
However,
it should be noted that the approximation within which one gets (\ref{m1}),
(\ref{g}) is well working for large enough  mass $M> 10M_{pl}$.
In principle, the correction terms $\sim ({M_{pl} \over M})^2$
to the classical laws could be expected. This needs somewhat more accurate
investigations.

\bigskip

\section{Conclusion}
\setcounter{equation}0

In this paper, we have analyzed the eternal black hole solution of the
two-dimensional
RST model giving us an example of exactly solvable one-loop effective theory.
The quantum-corrected  geometry of the black hole possesses  remarkable
properties.
Though the singularity is still present in the general solution, it becomes
milder than in the classical case. Moreover, the equations admit two copies of
the asymptotically flat black hole space-time defined on "different sides"
of the singularity. One of them (which is behind the singularity) does not have
a classical
analog. We propose that the complete space-time is a gluing of
both copies. It should be noted that a similar picture appears in different
quantum models \cite{15d}, \cite{PT}. The singularity is probably absent in the
complete quantum theory as it happens for the exact (non-perturbative) black
hole
background of the string theory \cite{PT}. But the global structure of the
black hole
space-time remains the same.

Generally,  quantum corrections are expected to change the thermodynamical
relations of a black hole. However, our consideration based on the exact
solution of
the RST model shows that this does not happen there. The mass, entropy and
temperature of the
quantum-corrected black hole are the same as in the classical case.
So, no quantum corrections! In principle, one can argue that this fact is a
feature
of this particular model but not a general property of the black hole physics.
This problem is worth studying on number of examples. On the other hand,
it follows from our consideration that the relation of entropy  and dates on
the horizon (in our case it is $\phi_h$) is really modified due to
quantum correction (see (\ref{4.13})). In terms of the 4D black hole physics
this can
be interpreted as the Bekenstein-Hawking entropy is really a more complicated
function of the area of the (quantum-corrected) horizon, $S_{BH}\sim {A_h \over
4}
-\kappa\ln A_h$, than
in the classics. The direct derivation of this result  in  four dimensions
needs
further investigation.

\bigskip

\begin{center}
{\bf Acknowledgments}
\end{center}
I  am grateful to  V.P.Frolov, W.Israel, R.C.Myers, for valuable
discussions.
This work was supported in part by  grant RFL300 of
the International Science Foundation and by  grant  N 94-02-03665-a
of RFFI.

\end{document}